\documentclass[letter]{aa} 
\usepackage{graphicx}
\usepackage{txfonts}
\begin{document}
   \title{Lithium depletion and the rotational history of exoplanet host stars}


   \author{J. Bouvier
          \inst{1}
          }

   \institute{Laboratoire d'Astrophysique, Observatoire de Grenoble,
              Universit\'e J. Fourier, CNRS, UMR~5571, BP~53, 38041
              Grenoble Cedex 9, France\\
              \email{jbouvier@obs.ujf-grenoble.fr}
             }

   \date{Received; accepted}

 
  \abstract
   {It has been reported that exoplanet host stars
are lithium depleted compared to solar-type stars without detected
massive planets. }
   {We investigate whether enhanced lithium depletion in 
exoplanet host stars may result from their rotational history.
}
   {We have developed rotational evolution models for slow and fast
solar-type rotators from the pre-main sequence (PMS) to the age of the
  Sun and compare them to the distribution of rotational periods
observed for solar-type stars between 1~Myr and 5~Gyr.}
   {We show that slow rotators develop a high degree of
differential rotation between the radiative core and the convective
envelope, while fast rotators evolve with little core-envelope
decoupling. We suggest that strong differential rotation at the
base of the convective envelope is responsible for enhanced lithium
depletion in slow rotators. }
   {We conclude that lithium-depleted exoplanet host stars were slow
rotators on the zero-age main sequence (ZAMS) and argue that slow
rotation results from a long lasting star-disk interaction during the
PMS. Altogether, this suggests that long-lived disks ($\geq$ 5~Myr)
may be a necessary condition for massive planet formation/migration.}

   \keywords{Stars: planetary systems : formation - Stars : rotation -
   Stars : abundances - Stars : pre-main sequence - Accretion,
   accretion disks - Hydrodynamics }

   \maketitle
%

\section{Introduction} 

As the discovery rate of exoplanets has steadily increased over the years,
now reaching nearly 300 detections, interest has grown in the
properties of their host stars (Santos et al. 2003; Kashyap et al. 2008). 
Israelian et al. (2004) report that solar-type stars with massive
planets are more lithium-depleted than their siblings without detected
massive planets. This result has recently been confirmed by Gonzalez
(2008). Lithium over-depletion in massive exoplanet hosts appears to
be a generic feature over a restricted T$_{eff}$ range from 5800 to
5950~K, independent of the planet's orbital properties. Metallicity
effects and/or the early accretion of planetesimals have been ruled
out as the origin of lithium over-depletion in exoplanet hosts (Castro
et al. 2008). Instead, it has been suggested that enhanced lithium burning
could stem from the tidal effect of the giant planet on the host
star, as it induces rotationally-driven mixing in the stellar interior
(Israelian et al. 2004). However, the mass of giant exoplanets is much
lower than the mass of the convective envelope of solar-type stars
($\geq$0.02$M_\odot$). In addition, enhanced lithium depletion is seen
for stars with giant planets that span a wide range of semi-major
axes, up to about 2~AU. Whether the tidal interaction between a
distant massive planet and the host star can be strong enough to
induce rotationally-driven mixing remains to be
investigated. 

Rotationally-driven mixing may instead be related to the rotational
history of the star (Takeda et al. 2007; Gonzalez 2008). We
revisit here the link between lithium depletion and the rotational
history of solar-type stars, and discuss its relationship with massive
planet formation. In Section 2, we present models for the rotational
evolution of solar-type stars from the early PMS to the age of the
Sun. We use these models to fit the most recent measurements of
rotational periods obtained for solar-type stars in star forming
regions and young open clusters. We discuss the properties of 2
extreme models: one reproduces the evolution of fast rotators over
time, and the other the evolution of slow rotators. We find that slow
rotators develop a high degree of differential rotation between the
inner radiative zone and the outer convective envelope, while fast
rotators evolve with little core-envelope decoupling. From these
results, we investigate in Section 3 the evolution of lithium
abundances in slow and fast rotators. We suggest that the large
velocity shear at the core/envelope interface in slow rotators leads
to more efficient Li-depletion. In Section 4, we relate the
Li-depletion pattern of massive exoplanet host stars to their
rotational history. We argue that both are dictated by their initial
angular momentum and the lifetime of their protoplanetary disk. 
We conclude that long-lived disk may be a necessary condition for
massive planet formation/migration around young solar-type stars.

\section{The rotational history of solar-type stars}

\subsection {Rotational evolution models} 

The rotational evolution of solar-mass stars depends on a number of
physical processes acting over the star's lifetime. The models
dicussed here were originally developed by Bouvier et al. (1997) and
Allain (1998), and we apply them to the most recent observational
constraints available today. The reader is referred to the original
papers for a complete description of the models, but we only briefly
recall the main assumptions and physical ingredients. The rotational
evolution of low-mass stars goes through 3 main stages: PMS evolution,
ZAMS approach, and MS relaxation. During the early PMS evolution, the
young star is magnetically coupled to its accretion disk. As long as
this interaction lasts, the star is prevented from spinning up as it
contracts and evolves at constant angular velocity. This
"disk-locking" assumption is supported by both observational evidence
(Rebull et al. 2006; Cieza \& Baliber 2007) and MHD models of the
star-disk interaction (e.g. Matt \& Pudritz 2005; Romanova et
al. 2007). The disk lifetime is a free parameter of the model, and it
dictates the early rotational evolution of the star. Then, as the disk
dissipates after a few Myr, the star spins up as it contracts and
becomes more centrally-condensed (due to the development of an inner
radiative zone) on its approach to the ZAMS\footnote{We used the
1~M$_\odot$ model from Baraffe et al. (1998) to follow the evolution
of the star's internal structure}. Depending on the initial velocity
and the disk lifetime, a wide range of rotation rates can be obtained
on the ZAMS (cf. Bouvier et al. 1997). The longer the disk lifetime
and the lower the initial velocity, the slower the rotation rate on
the ZAMS. In contrast, high initial velocities and/or short disk
lifetimes lead to fast rotation on the ZAMS. Finally, as the stellar
internal structure stabilizes at an age of about 40 Myr for a
solar-mass star, the braking by a magnetized wind becomes the dominant
process and effectively spins the star down on the early MS. As the
braking rate scales with surface velocity (see Kawaler 1988; Bouvier
et al. 1997), fast rotators are spun down more efficiently than slow
ones. As a result, and in spite of a large dispersion of rotation
rates on the ZAMS, solar-type stars converge towards uniformly slow
rotation by the age of the Sun. Indeed, after a few Gyr, the surface
rotational velocity of solar-type stars has lost all memory of the
past rotational history.

Internal differential rotation is an important additional parameter of
this model. The response of the stellar interior to angular momentum
loss at the stellar surface affects the evolution of the surface
rotation rate. We follow Allain (1998) in considering a radiative core
and a convective envelope that are each in rigid rotation, but whose
rotation rates may differ. We therefore introduce a coupling timescale
between the inner radiative zone and the outer convective envelope,
$\tau_c$, which measures the rate of angular momentum transfer between
the core and the envelope (MacGregor \& Brenner 1991). A short
coupling timescale corresponds to an efficient core-envelope angular
momentum transport and, as a consequence, little internal differential
rotation. In contrast, a long coupling timescale will lead to the
developement of a large rotational velocity gradient between the core
and the envelope. This model parameter is thus expected to be
essential for evaluating the amount of rotationally-induced mixing and
associated lithium depletion during the evolution of solar-type stars.

\subsection {Model fitting: slow and fast rotators}

   \begin{figure}
   \centering
  \includegraphics[width=0.5\textwidth]{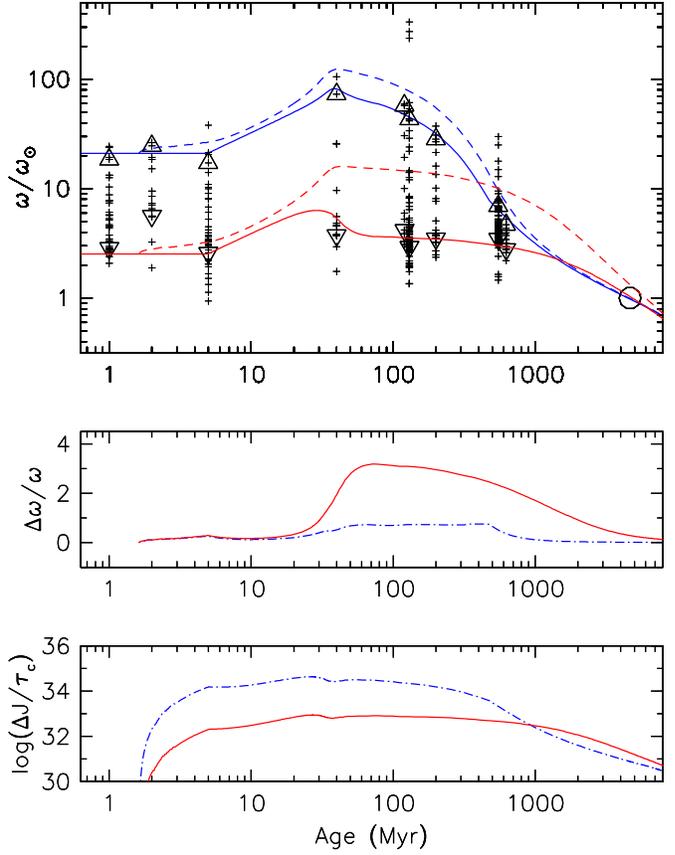}
   \caption{Rotational models for slow and fast rotators. {\it Data~:}
The 10th and 75th percentiles of the observed rotational period
distributions were converted to angular velocity and are plotted as
direct and inverted triangles as a function of time. Individual
measurements of rotational periods converted to angular velocities are
also shown in order to illustrate the statistical significance of the
various samples (see Table~1). {\it Models~:} The modeled evolution of
surface rotation for slow and fast rotators is shown by the solid
lines. For both models, the rotation of the radiative core is shown by
the dashed lines. With a core-envelope coupling timescale of only
10~Myr, little differential rotation develops in fast rotators. In
contrast, the 100~Myr core-envelope coupling timescale in slow
rotators leads to the development of a large velocity gradient at the
base of the convective zone. A disk lifetime of 5~Myr is assumed for
both models. {\it Lower panels :} The velocity shear at the base of
the convective zone ($\omega_{rad}-\omega_{conv})/\omega_{conv}$ and
the angular momentum transport rate $\Delta J/\tau_c$ ($g 
 cm^2  s^{-2}$) from the core to the envelope are shown for slow
(solid line) and fast (dotted-dashed line) rotators.}
              \label{model}%
    \end{figure}


\begin{table}
\caption{Observational datasets : 10th and 75th percentiles of the
rotational period distribution for 0.8-1.1~M$_\odot$ stars.}             
\label{rotobs}      
\centering                          
\begin{tabular}{l l l l l l}        
\hline\hline                 
Name & Age & P$_{10\%}$ & P$_{75\%}$ & N$_{star}$ & Ref.\\
& (Myr) & (d) & (d) \\
\hline
ONC & 1 & 1.4 & 8.9 & 35 & 1 \\
NGC~2264 & 2 & 1.0 & 4.5 & 16 & 2,3 \\
NGC~2362 & 5 & 1.5 & 9.8 & 40 & 4 \\
NGC~2547 & 40 & 0.35 & 6.8 & 15 & 5 \\
Pleiades & 120 & 0.44 & 6.1 & 11 & 6 \\
M50 & 130 & 0.59 & 8.7 & 74 & 7 \\
M34 & 200 & 0.89 & 7.2 & 22 & 8 \\
M37 & 550 & 3.6 & 7.3 & 145 & 9 \\
Hyades & 625 & 5.4 & 9.1 & 15 & 10 \\
\hline 
\ \\                                  
\multicolumn{6}{l}{(1) Herbst et al. (2002), (2) Lamm et al. (2005),}\\
\multicolumn{6}{l}{(3) Makidon et al. (2004), (4)  Irwin et  al. (2008b),}\\
\multicolumn{6}{l}{(5) Irwin et al. (2008a), (6) Compilation (see references  }\\
\multicolumn{6}{l}{in Irwin et al. 2008a), (7) Irwin (priv. comm.),}\\
\multicolumn{6}{l}{(8) Irwin et  al. (2006), (9) Hartman et al. (2008),}\\
\multicolumn{6}{l}{(10)  Radick et al. (1987, 1995)   }
\end{tabular}
\end{table}

The models are constrained by the rotation rates of solar-type stars
at various ages. In the last years, rotational periods have been
measured for hundreds of low-mass stars in molecular clouds (1-5 Myr)
and young open clusters (40-600 Myr), thus providing a tight
observational sampling of the rotational evolution of low-mass
stars. Table~1 lists the 10th and 75th percentiles of the derived
rotational period distributions for stars with a mass between 0.8 and
1.1~M$_\odot$ and belonging to various stellar clusters with an age
between 1~Myr and 625~Myr. These values were converted to angular
velocity ($\omega = 2\pi/P$), normalized to the Sun's
angular velocity, and are plotted as a function of time in
Figure~\ref{model}.

We aim here at reproducing the lower and upper envelopes of the
observed rotational distributions, in order to contrast the evolution
of slow and fast rotators and relate it to lithium depletion. A model
for fast rotators is compared to observations in
Fig.~\ref{model}. Starting from an initial period of 1.2~d, the star
remains coupled to its disk for 5~Myr, then spins up to a velocity of
about 160~km~s$^{-1}$ on the ZAMS, and is eventually spun down by a
magnetized wind on the MS to the Sun's velocity. This model fits
reasonably well the PMS spin up and the rapid MS spin down observed
for fast rotators between 5 and 500~Myr. To reach this 
agreement, the core-envelope coupling timescale has to be short,
$\tau_c \sim $ 10~Myr. A longer coupling timescale would lead to
envelope spin down before the star reaches the ZAMS, and a slower spin
down rate on the early MS, both of which would conflict with
observations. The tight coupling between the core and the envelope
implies that little differential rotation develops in fast rotators,
with the rotation of the core barely exceeding that of the envelope on
the early MS (see Fig.~\ref{model}).

Figure~\ref{model} also shows a model for slow rotators. The initial
period is 10~d, and the star-disk interaction lasts for 5~Myr in the
early PMS. As the star approaches the ZAMS, both the outer convective
envelope and the inner radiative core spin up. Once on the ZAMS, the
outer envelope is quickly braked, while the core remains in rapid
rotation. This behavior results from an assumed weak coupling between
the core and the envelope, with $\tau_c\sim$ 100~Myr. On the early MS,
the rapidly-rotating core transfers angular momentum back to the
envelope, which explains the nearly constant surface velocity over
several 100~Myr in spite of magnetic braking. A long coupling
timescale between the core and the envelope is thus required to
account for the observed rotational evolution of slow rotators. A
shorter coupling timescale would lead to higher ZAMS velocities and a
sharper spin down on the early MS, which both would conflict with
observations. A long $\tau_c$ in slow rotators implies inefficient
transport of internal angular momentum and results in a large velocity
gradient at the core-envelope boundary (see Fig.~\ref{model}).

The main difference between fast and slow rotator models is thus
twofold: the initial angular momentum and the level of core-envelope
decoupling. The initial angular momentum of fast rotators is 10 times
greater than that of slow ones. Fast rotators lose more angular
momentum over the course of their evolution than slow rotators, as
they both converge towards the Sun's velocity at 4.65~Gyr. The lower
panels in Fig.~\ref{model} show the magnitude of the rotational shear
at the base of the convective envelope for slow and fast rotators, and
the amount of angular momentum transported from the core to the
envelope. Differential rotation develops in slow rotators at the ZAMS
and remains strong during early MS evolution until the core and the
envelope eventually recouple after a few Gyr. Clearly, slow rotators
exhibit a much larger rotational shear at the base of their convective
envelope than do fast rotators during most of their evolution.
Conversely, angular momentum transport from the core to the envelope
is much more efficient in fast rotators than in slow ones, which leads
to very little differential rotation indeed.  Rotational shear and
angular momentum transport are both directly related to the
core-envelope coupling timescale and are independent of the assumed
disk lifetimes. Obviously, the different amounts of internal
differential rotation between slow and fast rotators is expected to
affect their lithium depletion pattern.

\section{Lithium depletion in slow and fast rotators}

Hydrodynamical instabilities acting on various timescales are
responsible for the so-called rotationally-driven mixing that
increases the efficiency of lithium burning in stellar interiors
(Talon 2008). The main sources of rotational mixing are shear-induced
turbulence and large-scale meridional circulation, both of which are
linked to internal differential rotation (Zahn 2007). These processes
transport angular momentum and chemical species, though not
necessarily at the same rate. Due to the complex physics involved, it
is by no means staightforward to predict from first principles which
instability dominates in solar-type stars over the course of their
evolution. Some guidance is provided by observations. A key result in
this respect is the observed relationship between Li abundances
and rotation in young stars. Soderblom et al. (1993) report that fast
solar-type stars in the Pleiades exhibit higher Li abundances
than slow rotators. Taken at face value, this result indicates that
Li depletion already takes place during the PMS, and is more
pronounced in slow rotators than in fast ones. Hence, rotationally-induced
mixing appears more efficient in slow rotators.

Going back to the models described above, this result suggests that
rotational mixing is primarily driven by the rotational shear at the
base of the convective envelope: slow rotators are observed to be
more Li-depleted on the ZAMS than fast rotators and the models
presented here indicate that they experience stronger core-envelope
decoupling. In contrast, fast rotators lose more angular momentum at
the stellar surface and also transfer angular momentum more
efficiently from the core to the envelope. Yet, they seem to preserve
most of their initial lithium during the PMS. Efficient angular
momentum transport is apparently not associated with a significant
mixing of chemical species in these stars. Mixing-free processes for
angular momentum transport have to be sought. Candidates include
magnetic field (Eggenberger et al. 2005) and internal gravity waves
(Talon et al. 2002). Whether magnetic fields, whether of fossil origin
or generated by a dynamo at the tachocline, can efficiently couple the
core to the envelope is still under debate (Brun \& Zahn 2006). And
while gravity waves seem capable of removing angular momentum from the
core, the efficiency of this process and the amount of associated
chemical mixing still remain to be fully explored (Charbonnel \& Talon
2005).

Comparison between rotational models and lithium measurements on the
ZAMS thus suggests that shear-induced turbulence associated to
core-envelope decoupling is the dominant source of rotationally-driven
lithium depletion in solar-type stars. Our finding that core-envelope
decoupling is stronger in slower rotators is at odds with most previous
theoretical studies. In Pinsonneault et al's (1989) models, for
instance, the rotational shear at the base of the convective envelope
scales with surface velocity. Hence, fast rotators are predicted to be
more heavily Li-depleted than slow ones. We show here instead
that, to be able to account for the observed rotational
evolution of solar-type stars from the PMS to the age of the Sun, one
has to assume that the velocity gradient at the base of the convective
envelope is larger in slow rotators than in fast ones (see also
Bouvier 1997). As a result, slow rotators on the ZAMS are expected to
be more efficiently Li-depleted than fast rotators, as observed.

\section{Lithium depletion and planet formation} 

We discussed above how different rotational histories may affect the
lithium depletion pattern of solar-type stars and thus lead to a {\it
dispersion\/} of lithium abundances in these stars at a given age and
mass. The scenario outlined above suggests that enhanced lithium
depletion is associated to low surface rotation on the ZAMS. Then,
that mature solar-type stars with massive exoplanets are Li-depleted
compared to similar stars with no planet detection seems to indicate
that massive exoplanet hosts had slow rotation rates on the ZAMS.

Why were massive exoplanet host stars slow rotators on the ZAMS?  Two
main parameters govern the rotation rate on the ZAMS : the initial
velocity and the disk lifetime. For any given disk lifetime, the lower
the initial velocity, the lower the velocity on the ZAMS. Conversely,
for any given initial velocity, the longer the disk lifetime, the lower
the velocity on the ZAMS. This is because the magnetic star-disk
interaction during the PMS is far more efficient than solar-type winds
in extracting angular momentum from the star (Bouvier 2007; Matt \&
Pudritz 2007). Disk lifetimes varying from star to star in the
range 1-10 Myr are required to account for the distribution of
rotational velocities on the ZAMS (Bouvier et al. 1997).
Statistically, however, the slowest rotators on the ZAMS are expected
to be the stars that initially had low rotation rates and the
longest-lived disks.

Long-lived disks may thus appear as a necessary condition for massive
planet formation and/or migration on a timescale $\geq$5~Myr. Long
lasting disks may indeed be the common origin for slow rotation on the
ZAMS, Li-depletion, and massive planet formation.  Interestingly
enough, the Sun is a massive exoplanet host. Even though the solar
system gaseous planets are located farther away from the Sun than
massive exoplanets are from their host stars, the Sun is strongly
Li-deficient. According to the scenario outlined above, the Sun would
thus have been a slow rotator on the ZAMS. More generally, another
intriguing result is the apparent bimodality of Li abundances in
mature solar-type stars (e.g., Pasquini et al. 1994, 2008). In the
framework of the model developed here, it is interesting to note that
the rotational distribution of young stars also appears bimodal (Choi
\& Herbst 1996). Whether the bimodal distributions of rotation and
lithium could be related through the rotational history of solar-type
stars remains to be addressed.

%
\section{Conclusions}

Based on what we currently know of the rotational properties of young
stars, of the lithium depletion process in stellar interiors, and of
the angular momentum evolution of solar-type stars, it seems likely
that the lithium-depleted content of massive exoplanet host stars is a
sequel to their specific rotational history. This history is
predominantly dictated by star-disk interaction during the PMS. Using
simple rotational models, it is suggested here that the long disk
lifetimes of initially slow rotators lead to the development of a
large velocity shear at the base of the convective zone. The velocity
gradient in turn triggers hydrodynamical instabilities responsible for
enhanced lithium burning on PMS and MS evolution
timescales. Rotationally-driven lithium depletion in exoplanet host
stars can thus be at least qualitatively accounted for by assuming PMS
disk lifetimes of 5-10 Myr. Such long-lived disks may be a necessary
condition for planet formation and/or migration around young
solar-type stars, at least for the class of giant exoplanets detected
so far (Ida \& Lin 2008).

Admittedly, the scenario outlined here is based on rotational models
that do not incorporate a detailed physical description of the
processes at work. The conclusion therefore remains qualitative and
has to be ascertained by the development of more sophisticated angular
momentum evolution models. Most important, the transport processes
that lead to a strong core-envelope coupling in fast rotators and a
weak one in slow rotators still have to be identified. While this goal
is well beyond the scope of the present study, we hope that the
simplified models presented here may provide useful guidelines for
further refinements.

\begin{acknowledgements}
It is a pleasure to acknowledge discussions of transport processes in
stellar interiors with C. Charbonnel, S. Mathis, A. Palacios,
S. Turck-Chieze, and J.-P. Zahn, during the 30th meeting of the French
Astronomical Society in Paris. I thank I. Baraffe for providing the
1~M$_\odot$ model, and J. Irwin for providing M50's rotational data
before publication.
\end{acknowledgements}

\end{document}